\documentclass{PoS}

\title{Observations of SN2011fe with INTEGRAL}

\ShortTitle{Gamma rays from SN2011fe}

\author{\speaker{J. Isern},$^a$ 
P. Jean,$^b$
 E. Bravo,$^c$ 
R. Diehl,$^d$
 J. Kn\"odlseder, $^b$
A. Domingo,$^e$
A. Hirschmann,$^a$
 P. Hoeflich,$^f$
 F. Lebrun,$^g$
 M. Renaud,$^h$
 S. Soldi,$^i$
 N. Elias--Rosa, $^a$
M. Hernanz,$^a$
 B. Kulebi,$^a$
 X. Zhang,$^d$
 C. Badenes,$^j$
 I. Dom\'{\i}nguez$^k$,
 D. Garcia-Senz,$^c$
 C. Jordi, $^l$
 G. Lichti,$^d$
G. Vedrenne$^b$ and
 P. Von Ballmoos$^b$
\thanks{Based on observations with \emph{INTEGRAL}, an ESA project with instruments and science data centre funded by ESA member states (specially the PI countries: Denmark, France, Germany, Italy, Switzerland, and Spain), the Czech Republic, and Poland and with the participation of Russia and USA.}\\
        \llap{$^a$}Institut de Ci\`encies de l'Espai (ICE-CSIC/IEEC), 
        \llap{$^b$}UPS-OMP; IRAP, Universit\'e de Toulouse, 
        \llap{$^c$}Dept. Fisica i Enginyeria Nuclear, Univ. Polit\`{e}cnica de Catalunya, 
        \llap{$^d$}Max -Planck-Institut for Extraterrestrial Physics, 
        \llap{$^e$}Centro de Astrobiolog\'{\i}a (CAB-CSIC/INTA), 
        \llap{$^f$}Physics Department, Florida State University, 
        \llap{$^g$}Astroparticle et Cosmologie (APC), CNRS-UMR 7164, Universit\'e de Paris 7 Denis   Diderot, 
        \llap{$^h$}Laboratoire Univers et Particules de Montpellier (LUPM), UMR 5299, Universit\'e de Montpellier II, 
       \llap{$^i$}AIM (UMR 7158 CEA/DSM-CNRS-Universit\'e Paris Diderot)
       Irfu/Service d'Astrophysique, 
       \llap{$^j$}Department of Physics and Astron omy \& Pittsburgh Particle Physics, Astrophysics and Cosmology Center (PITT-PACC), University of Pittsburgh, 
       \llap{$^k$}Universidad de Granada, 
       \llap{$^l$}Dept. d'Astronomia i Meteorologia, Institut de Ci\`encies del Cosmos (ICC), Universitat de Barcelona (IEEC-UB), 
        E-mail: \email{isern@ieec.cat}, pierre.jean@irap.omp.eu, eduardo.bravo@upc.edu,    rod@mpe.mpg.de, jknodlsede@irap.omp.eu, albert@cab.inta-csic.es, hirschmann@ieec.cat, pah@aastro.physics.fsu.edu, lebrun@apc.univ-paris7.fr, mrenaud@lupm.univ-montp2.fr, simona.soldi@cea.fr, nelias@ieec.cat, hernanz@ieec.cat, kulebi@ieec.cat, zhangx@mpe.mpg.de, inma@ugr.es, carme.jordi@ub.edu, grl@mpe.mpg.de, pvb@cesr.fr, vedrenne@cesr.fr}

\abstract{SN2011fe was detected by the Palomar Transient Factory on August 24th 2011 in M101 few hours after the explosion. From the early spectra it was immediately realized that it was a Type Ia supernova thus making this event the brightest one discovered in the last twenty years. In this paper the observations performed with the instruments on board of INTEGRAL (SPI, IBIS/ISGRI, JEM-X and OMC) before and after the maximum of the optical light as well as the interpretation in terms of the existing models of $\gamma$--ray emission from such kind of supernovae are reported. All INTEGRAL high-energy have only been able to provide upper limits to the expected emission due to the decay of $^{56}$Ni. These bounds allow to reject explosions involving a massive white dwarf in the sub--Chandrasekhar scenario. On the other hand, the optical light curve obtained with the OMC camera suggests that the event was produced by a delayed detonation of a CO white dwarf that produced $\sim 0.5$ M$\odot$  of $^{56}$Ni. In this particular case, INTEGRAL would have only been able to detect the early $\gamma$--ray emission if the supernova had occurred at a distance  of 2 -3 Mpc, although the brightest event could be visible up to distances larger by a factor two. The observation of a statistically representative sample of SNIa demands sensitivities of $\sim 10^{-7}$ cm$^{-2}$s$^{-1}$keV$^{-1}$.}

\FullConference{"An INTEGRAL view of the high-energy sky (the first 10 years)"
9th INTEGRAL Workshop and celebration of the 10th anniversary of the launch,\\
		October 15-19, 2012\\
		Bibliotheque Nationale de France, Paris, France}

\begin{document}

\section{Introduction}

Type Ia supernovae (SNIa) are characterized by the lack of H--lines and the presence of Si II--lines in their optical spectra during the maximum of light as well as by the presence of Fe emission features during the nebular phase. Their optical light curve displays a sudden rise to the maximum followed by a rapid decay of $\sim 3$~mag in one month and by a slowly-- fading tail.  A noticeable property is the spectrophotometric homogeneity of the different outbursts. Furthermore, they appear in all kinds of galaxies including ellipticals. These properties point to an exploding object that is compact, free of hydrogen, that can be activated on short and long time scales, and is able to synthesize enough $^{56}$Ni to power the light curve. The most obvious candidate is a C/O white dwarf (WD) near the Chandrasekhar's limit in a close binary system that explodes as a consequence of mass accretion \cite{hoyl60}. 

Despite their homogeneity, SNIa display some differences when observed in detail. Now it is
known that there is  a group of  SNIa with light  curves showing very  bright and
broad peaks,  the SN1991T  class, that represents  9\% of  all the
events. There  is another  group with a much dimmer and narrower peak and that
lacks the  characteristic secondary peak in the infrared, the SN1991bg  class, that
represents 15\% of all the  events. To these categories it has been
recently added a  new one that contains very  peculiar supernovae, the
SN2002cx or SNIax class,  representing $\sim 5$\%  of the  total. These
supernovae are characterized by high ionization spectral features in
the pre-maximum, like the SN1991T  class, a very low luminosity,
and  the lack  of a secondary  maximum in the infrared,  like the  SN1991bg class.   The
remaining ones, which amount  to $\sim 70\%$, present normal behaviors and are  known as 
\emph{Branch-normal} \cite{li11a}. However, even the  normal ones
are  not completely  homogeneous  and show  different luminosities  at
maximum and light curves with different decline rates \cite{li11b}. This variety has recently increased with the discovery of SN2001ay, which is characterized by a fast rise and a very slow decline \cite{baro12}. This diversity strongly suggests that different scenarios and burning 
mechanisms could be operating in the explosion.

In one dimension models, the explosion mechanism can be classified as \cite{hoef96,hill00}: the pure detonation model (DET), the pure deflagration model (DEF), the delayed detonation model (DDT), and the pulsating delayed detonation model (PDD).  An additional class  are the so called Sub--Chandrasekhar's (SCh) models in which a detonation triggered by the ignition of He near the base  of a freshly accreted helium layer completely burns the white dwarf. At present, there is no basic argument to reject any of them, except the DET ones that are incompatible with the properties of 
the spectrum of SNIa at maximum light. Present observations also pose severe constraints to 
the total amount of $^{56}$Ni that can be produced by the He--layer in SCh models. The equivalent models in three dimensions also exist, but with a larger variety of possibilities (see for instance \cite{brav09}).

According to the nature of the binary, progenitors can be classified as single degenerate (SD) if the companion is a normal star \cite{whel73} or double degenerate (DD) if the companion is a white dwarf \cite{webb84,iben85}. The distinction among them is important in order to interpret the observations since, depending on the case, the white dwarf can ignite below, near or above the Chandrasekhar's mass and the total mass ejected as well as the mass of $^{56}$Ni synthesized can be different. It is not known if both scenarios can coexist or just one is enough to account for the supernova variety. Observations of the stellar content in the interior of known SNIa remnants point towards one possible SD candidate in the case of Tycho Supernova \cite{ruiz04}, two almost certain DD candidates, SNR0509-67.5 and SNR0519-69.0 \cite{scha12,edwa12}, and the new evidence that there is not a surviving companion in SN1006 \cite{gonz12}.

The amount and distribution of the radioactive material produced in the explosion strongly depend on how the ignition starts and how the nuclear flame propagates \cite{gome98,iser08}. Therefore, the detection of $\gamma$--rays from supernovae could provide important insight on the nature of the progenitor and especially on the explosion mechanism. The advantages of using $\gamma$--rays for diagnostic purposes rely on their ability to distinguish among different isotopes and on the relative simplicity of their transport modelling. 
In the case of close enough outbursts, less than $\sim 1$~Mpc, it would be possible to obtain high quality $\gamma$--ray spectra that could allow to perform detailed comparisons with theoretical predictions. However, when more realistic distances are considered, the information provided by observations decreases drastically and only some outstanding features, like the intensity of the lines and of the continuum and the line profiles, have a chance to be detected \cite{gome98}. Because of the scarcity of close enough events up to now it has only been possible to place upper limits to the SN1991T \cite{lich94} and SN1998bu \cite{geor02} events.  

\section{INTEGRAL observations of SN 2011fe}

SN 2011fe (RA = 14:03:05.81, Dec = +54:16:25.4; J2000) was discovered in M101 on August 24th, 2011, $\sim 1$ day after the explosion \cite{nuge11}. The absence of hydrogen and helium, coupled with the presence of silicon in the spectrum clearly indicated it was a SNIa. The distance of M101, 6.4 Mpc, is slightly less than the maximum distance at which current gamma-ray instruments should be able to detect an intrinsically luminous SNIa. The closeness of SN2011fe has made it possible to obtain the tightest constraints on the supernova and its progenitor system, leaving only either DD or a few cases of SD as possible progenitor systems of this supernova. 

\begin{table}
\caption{Upper-limit of the flux in selected spectral regions for SPI (2$\sigma$), JEM--X (2$\sigma$), and IBIS/ISGRI (3$\sigma$) for the entire pre and post--maximum observation periods (from days 5 to 18 and 45 to 88 after the explosion respectively).}             
\label{tab1}      
\centering                  
\begin{tabular}{ccc}        
\hline\hline                 
\multicolumn{3}{c}{Early period}  \\
\hline
Energy band  & Upper-limit flux                       & Instrument  \\ 
 (keV)            & (photons s$^{-1}$ cm$^{-2}$) &                     \\  
 \hline                        
3 - 10           &  $5.0 \times 10^{-4}$  & JEM-X       \\
10 - 25         &  $4.0 \times 10^{-4}$  & JEM-X       \\
3 - 25           &  $1.0 \times 10^{-3}$  & JEM-X       \\ 
60 - 172       &  $1.5 \times 10^{-4}$  & IBIS/ISGRI \\
90 - 172       &  $1.1 \times 10^{-4}$  & IBIS/ISGRI \\
150 - 172     &  $7.1 \times 10^{-5}$  & IBIS/ISGRI \\    
160 - 166     &  $7.5 \times 10^{-5}$  & SPI            \\     
140 - 175     &  $2.3 \times 10^{-4}$  & SPI            \\     
814 - 846     &  $2.3 \times 10^{-4}$  & SPI            \\     
800 - 900     &  $3.5 \times 10^{-4}$  & SPI            \\     
\hline \hline  
\multicolumn{3}{c}{Late period}\\
\hline
Energy band  & Upper-limit flux                        & Instrument  \\ 
 (keV)            & (photons s$^{-1}$ cm$^{-2}$)  &                    \\  
 \hline                        
 505 - 525 & $1.1 \times 10^{-4}$& SPI \\
 830 - 875 & $1.4 \times 10^{-4}$&SPI \\
 835 - 870 & $1.2 \times 10^{-4}$& SPI \\ 
1215-1275& $1.2 \times 10^{-4}$&SPI \\
1220-1270& $1.1 \times 10^{-4}$&SPI \\
1225-1265& $1.0 \times 10^{-4}$&SPI \\    
\hline                                      
\end{tabular}
\end{table}

\emph{INTEGRAL}  observed this supernova with its four instruments (SPI, ISGRI/IBIS, JEM--X,  and OMC) before the peak of the light curve, from August 29th to September 12th 2011 or, equivalently,  from day $\sim 5$ to day $\sim 18$ after the explosion, and after the peak, from October 7th to November 19th, 2011 or, equivalently, from day $\sim 45$ to day $\sim 88$. 
The early observations were essentially constrained by the Sun, that  prevented the observation just after the optical maximum where the $^{56}$Ni lines are expected to peak. The results of JEM-X, ISGRI and SPI observations, displayed in Table \ref{tab1}, were only upper limits.

\begin{figure}
\centering
\includegraphics[width=.6\textwidth]{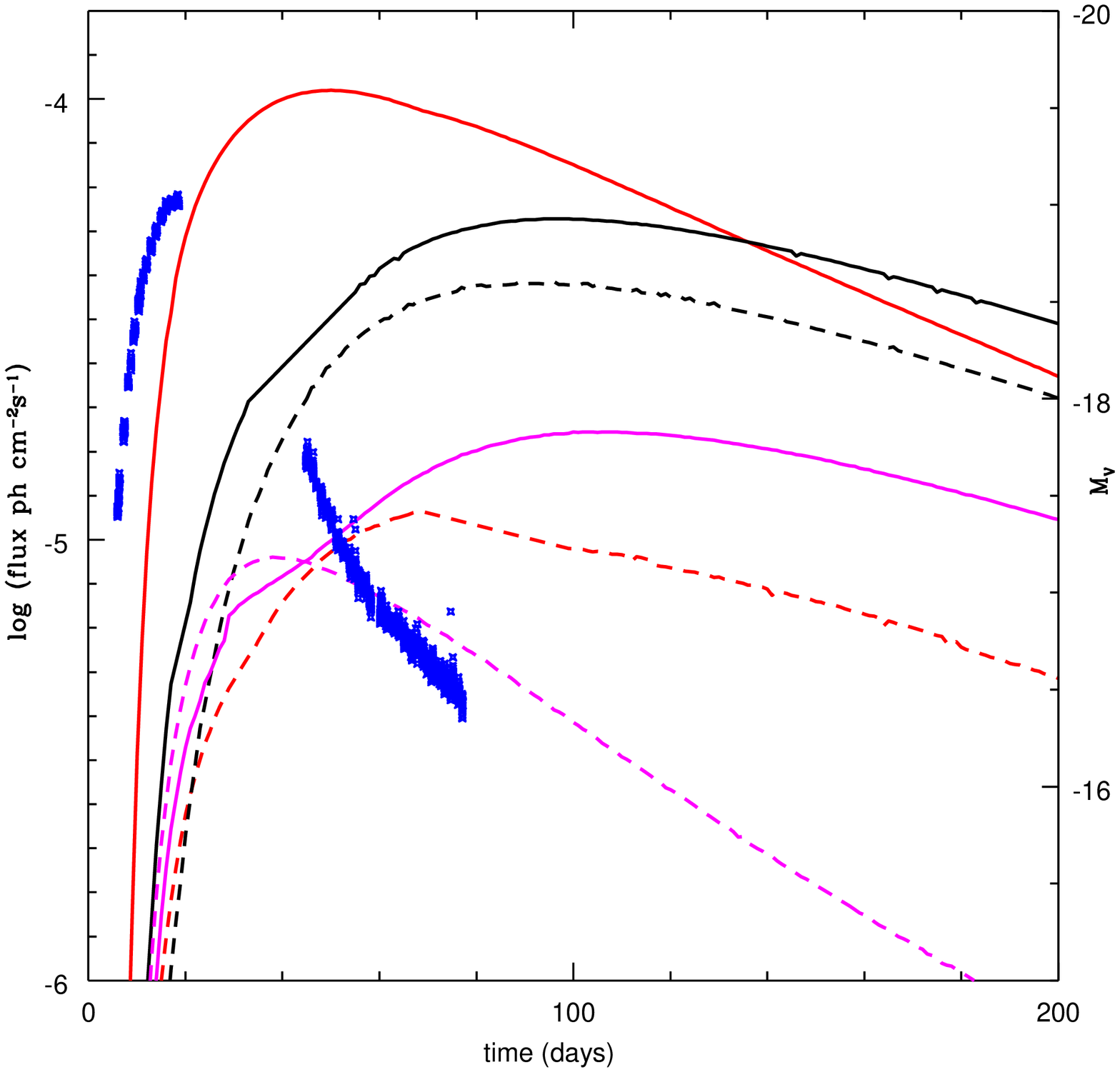}
\caption{SN2011fe predicted evolution of the $^{56}$Ni 158, 812 keV features (dashed and continous magenta lines, respectively), the $^{56}$Co 847, 1238 keV features (continous and dashed black lines) and the 511 keV annihilation line as well as the band, 200-540 keV (dashed and continuous red line). Blue dots represent the evolution of the SN2011fe visual magnitude as obtained with the OMC camera of INTEGRAL.}
\label{figlc}
\end{figure}

Figure \ref{figlc} displays the light curve in the optical V--band obtained with the OMC. The magnitude at maximum was $M_V = -19.04$, thus indicating that SN2011fe was a slightly dim average SNIa. This light curve is well fitted with a delayed detonation model of a Chandrasekhar mass white dwarf igniting at $\rho_c = 2\times 10^9$ g/cm$^3$ and making the transition from deflagration to detonation at $\rho_{tr} = 2.2 \times 10^7$ g/cm$^3$. The corresponding decline parameter in the blue was $\Delta m_{\rm 15} =1.2 \pm 0.2$ in agreement with the Phillips relationship. The total amount of $^{56}$Ni synthesized in this way was in the range of $\sim 0.51-0.55$ M$_\odot$ when the uncertainties in the values of the interstellar extinction are taken into account.

\begin{figure}
\centering
\includegraphics[width=.8\textwidth]{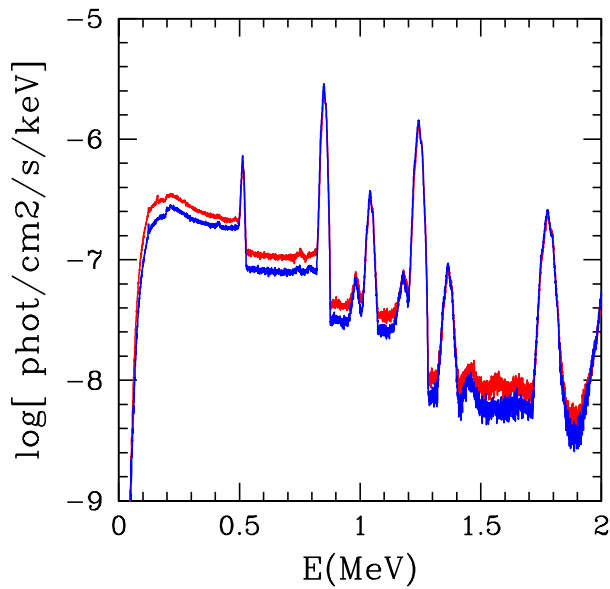}
\caption{Predicted spectra from the DDTe model at 70 (red line) and 90 (blue line) days after the explosion.}
\label{figspc}
\end{figure}

The expected gamma ray  emission of the above model has been obtained with the code described in \cite{gome98}, which was successfully cross--checked with other independent codes \cite{miln04}. Figure \ref{figlc} displays the evolution with time of several important gamma ray features. The 200-540 keV band contains almost all the annihilation photons and it is the brightest feature. The $^{56}$Ni lines are characterized by their sharp rise and their relatively rapid decline as a consequence of the rapid expansion of the debris, with the corresponding increase of the transparency, and the relatively short lifetime of this isotope. The slow decline of the 812 keV line is a consequence of the overlapping with the growing 847 keV $^{56}$Co line. The $^{56}$Co lines have a more gentle growing and a slow decay as a consequence of the larger lifetime of the isotope and the increasing transparency of the envelope. Figure \ref{figspc} displays the spectra 70 and 90 days after the explosion. The main characteristics, valid for almost every epoch, is the extremeley large width of the lines ($\Delta E/ E \gtrsim 5\%$) as a consequence of Doppler and Compton broadening.

\section{Discussion and conclusions}
Besides the relative weakness of the source, the width of the lines and the rapid variation of their intensity are responsible for the non detection of SN2011fe by \emph{INTEGRAL}. In the limit of weak signals \cite{jean96}, the significance is given by:

\begin{equation}
n_{\sigma} = \frac{{A_{eff} \int\limits_{t_i  - \Delta t}^{t_i } {\varphi \left( t \right)dt} }}{{\sqrt {bV\Delta E \Delta t} }} 
\label{nsigma}
\end{equation}

\noindent where $\Delta t$ is the effective observation time, $A_{eff}$ is the effective area at the
corresponding energies, $\varphi$ is the flux  (cm$^{-2}$s$^{-1}$) 
in the energy band $\Delta E$,  V is the volume of the detector and b is the 
specific noise rate (cm$^{-3}$s$^{-1}$keV$^{-1}$), where it has been assumed 
that it is weakly dependent of the energy and time in the interval of interest. It is evident from Eq. \ref{nsigma} that if the photons produced by a nuclear transition are distributed over an energy band $\Delta E$, the noise will grow as $ \propto \sqrt{\Delta E} $ and the significance of the signal will decrease as compared with the thin line case. Table \ref{tab2} displays the width that optimizes the signal to noise ratio at maximum in the DDTe case.

It is also evident from Eq. \ref{nsigma} that if the flux changes with time, the significance will also change since it will be a function of the total number of photons detected during the observation time. For instance, assume that the flux follows an exponential law like $ \varphi \left( t \right) = \varphi _0 e^{\alpha t} $. The significance of the signal obtained integrating during the time interval $\left( {t_i  - \Delta t,t_i } \right)$ is

\begin{equation} 
 n = \frac{{A_{eff} \varphi \left( {t_i } \right)}}{{\sqrt {\alpha bV\Delta E} }}\frac{{1 - e^{ -
\alpha \Delta t} }}{{\sqrt {\alpha \Delta t} }} 
\end{equation}

 \noindent For $\alpha \Delta t <  < 1$, it behaves as $n \propto \sqrt{\Delta t}$ and has a
maximum at $ \alpha \Delta t = 1 .26$ in the general case. This dependence clearly shows the convenience to take a value of $\Delta t$ that maximizes the signal/noise ratio. Unfortunately, since the value of $\alpha$ is not known \emph{a priori} the optimal observing time is not known in advance \cite{iser13}. 

\begin{table}
\caption{Width of the lines that optimizes the S/N ratio at the maximum of their intensity in the case of the DDTe model}             
\label{tab2}      
\centering                  
\begin{tabular}{cc}        
\hline\hline                 
Energy (keV)  &  $\Delta E$ (keV)  \\ 
 \hline                        
158            &  20      \\
511            &  30      \\
511 (band) &  340    \\ 
812            &  35      \\
847            &  27      \\
1238         &  40       \\        
\hline                                      
\end{tabular}
\end{table}

For instance, if we assume that the DDTe model (see Fig. \ref{figlc}) is representative of SN 2011fe, we see that at the beginning, when the emission is dominated by the $^{56}$Ni and ejecta are still opaque, there is a rapid growing of the intensity of the lines and only the last five days of the first period of observation are useful. On the contrary, when lines are dominated by the $^{56}$Co lines and matter is already transparent, the temporal behavior is more gentle, the above restriction does not apply and it is possible to follow a cumulative strategy of observation. Therefore, estimating in advance the significance of an observation demands a previous knowledge of the evolution of the shape and intensity of the line.

It is clear from the previous discussion that the detectability of a supernova not only depends on the distance but also on the subtype since the intensity of the radioactive lines is a function of the total amount of $^{56}$Ni synthesized and this quantity goes from $\sim 1$ M$_\odot$ in the case of a superluminous SNIa to $\sim 0.05$ M$_\odot$ in the case of a dim SNIax. Neglecting the last family and only taking the well classified events that appear in the catalog of the Sternberg Astronomical Institute, it turns out that to detect $\sim 6$ bright events per year it is necessary to achieve a sensitivity of the order of $\sim 10^{-7}$ cm$^{-2}$s$^{-1}$keV$^{-1}$. Of course, these figures are approximate, but they represent the frontier that allows to set up a systematic program of observation of SNIa or to consider these events as serendipitous ToO.

\acknowledgments
This  work has been  supported by the  MINECO-FEDER grants AYA2011-24704/ESP, AYA2011-24780/ESP,  AYA2009-14648-C02-01, CONSOLIDER CSC2007-00050,  by the ESF EUROCORES Program EuroGENESIS  (MINECO grants  EUI2009-04170), by the grant 2009SGR315 of the Generalitat de Catalunya. In parts, this work has also been supported by the NSF grants AST-0708855 and AST-1008962 to PAH.

The INTEGRAL SPI project has been completed under the responsibility and leadership of CNES.
We also acknowledge the INTEGRAL Project Scientist Chris Winkler (ESA, ESTEC) and the INTEGRAL personnel for their support to these observations.


\begin{thebibliography}{99}
\bibitem{hoyl60} 
F.~Hoyle, W.A.~Fowler, \emph{Nucleosynthesis in Supernovae}, \emph{ApJ} {\bf 132} (1960) 565
\bibitem{li11a}
W.~Li, et al.  \emph{Nearby supernova rates from the Lick Observatory Supernova Search - II. The observed luminosity functions and fractions of supernovae in a complete sample}, \emph{MNRAS} {\bf 412} (2011) 1441
\bibitem{li11b}
W.~Li, et al. \emph{Nearby supernova rates from the Lick Observatory Supernova Search - III. The rate-size relation, and the rates as a function of galaxy Hubble type and colour}, \emph{MNRAS} {\bf 412} (2011) 1473
\bibitem{baro12}
E.~Baron, et al. \emph{A Physical Model for SN 2001ay, a Normal, Bright, Extremely Slow Declining Type Ia Supernova}, \emph{ApJ} {\bf 753} (2012) 105
\bibitem{hoef96}
P.~Hoeflich, et al., \emph{Maximum Brightness and Postmaximum Decline of Light Curves of Type IA Supernovae: A Comparison of Theory and Observations}, \emph{ApJ} {bf 472} (1996) L81
\bibitem{hill00}
W.~Hillebrandt,  J.C.~Niemeyer, \emph{ARAA}, \emph{Type IA Supernova Explosion Models} {\bf 38} (2000) 191
\bibitem{brav09}
E.~Bravo, D.~Garcia-Senz, \emph{Pulsating reverse detonation models of SNIa. I Detonation ignition}, \emph{ApJ} {\bf 695} (2009) 1244
\bibitem{whel73}
J.~Whelan, I.~Iben, \emph{Binaries and Supernovae of Type I}, \emph{ApJ} {\bf 186} (1973) 1007
\bibitem{webb84}
R.F.~Webbink, \emph{Double white dwarfs as progenitors of R Coronae Borealis stars and Type I supernovae}, \emph{ApJ} {\bf 277} (1984) 355
\bibitem{iben85}
I.~Iben, A.V.~Tutukov, \emph{On the evolution of close binaries with components of initial mass between 3 solar masses and 12 solar masses}, \emph{ApJ} {\bf 58} (1985) 661 
\bibitem{ruiz04}
P.~Ruiz-Lapuente, et al., \emph{The binary progenitor of Tycho Brahe's 1572 supernova}, \emph{Nature} {\bf 431} (2004) 1069
\bibitem{scha12}
B.F.~Schaefer, A.~Pagnotta, \emph{An absence of ex-companion stars in the type Ia supernova remnant SNR 0509-67.5}, \emph{Nature} {\bf 481} (2012) 164
\bibitem{edwa12}
Z.I.~Edwards, A.~Pagnotta, B.E.~Schaefer, \emph{The Progenitor of the Type Ia Supernova that Created SNR 0519-69.0 in the Large Magellanic Cloud}, \emph{ApJ} {\bf 747} (2012) L19
\bibitem{gonz12}
J.I.~Gonzalez-Hernandez, et al., \emph{No surviving evolved companions of the progenitor of SN1006}, \emph{Nature} {\bf 489} (2012) 533
\bibitem{gome98}
J.{Gomez-Gomar, J.~Isern, P.~Jean, \emph{Prospects for Type IA supernova explosion mechanism identification with gamma rays}, \emph{MNRAS} {\bf 295} (1998) 1
\bibitem{iser08}
J.~Isern, E.~Bravo, A.~Hischmann, \emph{Detection and interpretation of {$\gamma$}-ray emission from SNIa}, \emph{New Astr. Rev.} {\bf 52} (2008) 377
\bibitem{lich94}
G.G.~Lichti, et al., \emph{COMPTEL upper limits on gamma-ray line emission from Supernova 1991T}, \emph{AA} {\bf 292} (1994) 569
\bibitem{geor02}
R. Georgii, et al., \emph{COMPTEL upper limits for the $^{56}$Co gamma -ray emission from SN1998bu}, \emph{AA} {\bf 394} (2002) 517
\bibitem{nuge11}
P.E.~Nugent, et al. , \emph{Supernova SN 2011fe from an exploding carbon-oxygen white dwarf star}, \emph{Nature} {\bf 480} (2011) 344
\bibitem{miln04}
P.A.~Milne, et al., \emph{Unified One-Dimensional Simulations of Gamma-Ray Line Emission from Type Ia Supernovae}, \emph{ApJ} {\bf 613} (2004) 1101
\bibitem{jean96}
P.~Jean, \emph{Etudes des performances et modelisation de spectrometres gamma pour l'Astrophysique Nucleaire}, PhD Thesis, Universite Paul Sabatier, Toulouse (1996).
\bibitem{iser13}
J.~Isern, et al., \emph{Observation of SN2011fe with INTEGRAL. I. Pre-maximum phase}, \emph{AA} (2013), in press (arxiv 1302.3381)
}
\end{thebibliography}
\end{document}